\documentclass{article}
\usepackage{times}
\usepackage{amsfonts}
\usepackage{graphicx}
\begin{document}
\hyphenation{di-men-sion-al}
\noindent
{\Large AMPLITUDE, PHASE, AND COMPLEX ANALYTICITY}\\

\vskip1cm
\noindent
{\bf D. Cabrera},$^{a}$ {\bf P. Fern\'andez de C\'ordoba}$^{b}$ and {\bf  J.M. Isidro}$^{c}$\\
Instituto Universitario de Matem\'atica Pura y Aplicada,\\ Universidad Polit\'ecnica de Valencia, Valencia 46022, Spain\\
${}^{a}${\tt dacabur@upvnet.upv.es}, ${}^{b}${\tt pfernandez@mat.upv.es}\\
${}^{c}${\tt joissan@mat.upv.es}\\
%\vskip.5cm
%\noindent
%\today
\vskip.5cm

\noindent
{\bf Abstract} Expressing the Schroedinger Lagrangian ${\cal L}$ in terms of the quantum wavefunction $\psi=\exp(S+{\rm i}I)$ yields the conserved Noether current ${\bf J}=\exp(2S)\nabla I$. When $\psi$ is a stationary state, the divergence of ${\bf J}$ vanishes. One can exchange $S$ with $I$ to obtain a new Lagrangian $\tilde{\cal L}$ and a new Noether current $\tilde{\bf J}=\exp(2I)\nabla S$, conserved under the equations of motion of $\tilde{\cal L}$. However this new current $\tilde{\bf J}$ is generally not conserved under the equations of motion of the original Lagrangian ${\cal L}$. We analyse the role played by $\tilde{\bf J}$ in the case when classical configuration space is a complex manifold, and relate its nonvanishing divergence to the inexistence of complex--analytic wavefunctions in the quantum theory described by ${\cal L}$.
%\tableofcontents
\section{Introduction}\label{sekt1}

Madelung, Brillouin, Kramers and Wentzel with their WKB approximation, and later Bohm,  all pioneered the factorisation of the complex wavefunction $\psi$ into amplitude and phase (see the book \cite{HOLLAND} for a full account). Invoking the correspondence principle, this factorisation expresses the phase of $\psi$ as the (complex) exponential of the classical action ${\cal I}$. For the amplitude of $\psi$ one invokes Boltzmann's principle and Born's rule in order to write it as the (real) exponential of the entropy ${\cal S}$ \cite{ENTROPY}. Altogether one writes the wavefunction as
\begin{equation}
\psi=\exp\left(\frac{{\cal S}}{2k_B}+{\rm i}\frac{{\cal I}}{\hbar}\right),
\label{ansaz}
\end{equation}
where $k_B$ is Boltzmann's constant.\footnote{That Boltzmann's constant $k_B$ qualifies as a quantum of entropy does not seem to have been widely recognised in the literature; see however ref. \cite{LANDAUER}.} Obviously the factorisation (\ref{ansaz}) breaks down at the zeroes of $\psi$, where one formally sets ${\cal S}=-\infty$. It is convenient to introduce the dimensionless entropy $S$ and the dimensionless mechanical action $I$,
\begin{equation}
S:=\frac{{\cal S}}{2k_B}, \qquad I:=\frac{{\cal I}}{\hbar},
\label{foor}
\end{equation}
in order to write more neatly
\begin{equation}
\psi=\exp\left(S+{\rm i}I\right).
\label{limppo}
\end{equation}
Then the quantum mechanics of $\psi$ can be very conveniently pictured as the fluid mechanics of a quantum probability fluid, the velocity field ${\bf v}$ being given by
\begin{equation}
{\bf v}=\frac{\hbar}{m}\nabla {I}.
\label{atled}
\end{equation}

Using the decompostion (\ref{limppo}), in ref. \cite{ENTROPY}  we have analysed the properties of {\it nonstationary}\/ quantum states. In the present letter we will analyse {\it stationary}\/ states instead. We will establish that {\it when classical configuration space is $\mathbb{R}^{2k}$, quantum effects cause a certain current $\tilde{\bf J}$ (to be defined in Eq. (\ref{jota}) below) to develop a nonvanishing value of\/ $\nabla\cdot\tilde{\bf J}$}\/.  In the stationary regime, a nonvanishing value of $\nabla\cdot\tilde{\bf J}$ indicates the breakdown of a conservation law. Specifically, we will establish conditions under which $\nabla\cdot\tilde{\bf J}\neq 0$ will imply {\it the impossibility of having complex--analytic wavefunctions on $\mathbb{R}^{2k}$}\/, although the latter qualifies as a complex--analytic manifold. Analytic wavefunctions are common in the theory of coherent states \cite{PERELOMOV}, but they are defined on phase space instead.

We will see that the quantum effect responsible for the lack of complex analyticity in the quantum theory is the appearance of a natural length scale $l$ associated with a quantum particle, namely the de Broglie wavelength $\lambda_B=\hbar/(mv)$.  One the contrary, no natural length scale exists in classical mechanics. This does not imply that classical mechanics can always be endowed with a complex--analytic structure.  A necessary condition for complex analyticity is that the dimension of configuration space be even, so odd--dimensional spaces are ruled out already from the start. 

For the rest of this letter we will consider a quantum particle with $\mathbb{R}^{2k}$ as its classical configuration space. Then the stationary wavefunction $\psi$ will be a complex--valued function $\psi:D\subset\mathbb{R}^{2k}\rightarrow\mathbb{C}$ depending on $2k$ real coordinates $x_j$. Let $u=u(x_j)$ and $v=v(x_j)$ be real--valued functions on the domain $D$. We recall that the Cauchy--Riemann equations for the complex function $u+{\rm i}v$ are equivalent to the orthogonality condition $\nabla u\cdot\nabla v=0$, where $\nabla=(\partial_{x_1}, \ldots,\partial_{x_{2k}})$.

\section{Analyticity and the equations of motion}\label{sekt2}

We will describe the quantum motion of a particle of mass $m$ under the stationary, external potential $V=V(x_j)$ by means of a fluid flow in a domain $D$ within configuration space $\mathbb{R}^{2k}$. It is convenient to start with the Schroedinger Lagrangian,
\begin{equation}
{\cal L}={\rm i}\hbar\psi^*\partial_t\psi-\frac{\hbar^2}{2m}\nabla\psi^*\nabla\psi-V\psi^*\psi,
\label{shrodi}
\end{equation}
and perform a canonical analysis in terms of the variables $S$ and $I$, as in Eq. (\ref{limppo}). Substituting $\psi=\exp(S+{\rm i}I)$ we find 
\begin{equation}
{\cal L}=\exp(2S)\left\{{\rm i}\hbar(\partial_tS+{\rm i}\partial_tI)-\frac{\hbar^2}{2m}\left[(\nabla S)^2+(\nabla I)^2\right]-V\right\}.
\label{eleeseu}
\end{equation}
This Lagrangian is manifestly invariant under the transformations $I\rightarrow I+\alpha$, where $\alpha\in\mathbb{R}$. These transformations induce rigid phase rotations $\psi\rightarrow\exp({\rm i}\alpha)\psi$ of the wavefunction. The corresponding conserved Noether current is the probability density current,
\begin{equation}
{\bf J}=\frac{\hbar}{m}\exp(2S)\nabla I=\rho{\bf v},
\label{novajota}
\end{equation}
where the dimensionless density function is $\rho=\exp(2S)$, and the velocity field ${\bf v}$ is given in Eq. (\ref{atled}).

{}From the Lagrangian (\ref{eleeseu}) one derives the equations of motion for $S$ and $I$. The former is also known as the quantum Hamilton--Jacobi equation,
\begin{equation}
\hbar\frac{\partial I}{\partial t}+\frac{\hbar^2}{2m}\left(\nabla I\right)^2+V+{\cal U}=0,
\label{qhj}
\end{equation}
where ${\cal U}$ denotes the quantum potential,
\begin{equation}
{\cal U}:=-\frac{\hbar^2}{2m}\left[\left(\nabla S\right)^2+\nabla^2 S\right].
\label{cuantist}
\end{equation}
The equation of motion for $I$ is the continuity equation for the quantum probability fluid,
\begin{equation}
\frac{\partial S}{\partial t} + \frac{\hbar}{m}\nabla S\cdot \nabla I+\frac{\hbar}{2m}\nabla^2 I=0.
\label{ryman}
\end{equation}
Stationarity means $\partial S/\partial t=0$ and $\partial I/\partial t=-E$, thus Eqs.  (\ref{qhj}) and  (\ref{ryman}) respectively become
\begin{equation}
\frac{\hbar^2}{2m}\left(\nabla I\right)^2+V+{\cal U}=E
\label{estaz}
\end{equation}
and
\begin{equation}
\nabla S\cdot \nabla I=-\frac{1}{2}\nabla^2 I.
\label{ryhbfb}
\end{equation}
If we now compute the divergence of the probability density current (\ref{novajota}) we find that, by Eq. (\ref{ryhbfb}), it vanishes identically as had to be the case:
\begin{equation}
\nabla\cdot{\bf J}=\frac{\hbar}{m}\exp(2S)\left(2\nabla S\cdot\nabla I+\nabla^2 I\right)=0.
\label{bans}
\end{equation}
The conservation law (\ref{bans}) implies that the functions $S$ and $I$ can be arranged into an {\it analytic}\/ function 
\begin{equation}
g:=S+{\rm i}I
\label{geje}
\end{equation}
if and only if $I$ is harmonic. Let us summarise:

{\bf Property 1} {\sl In the quantum mechanics of a particle in $D\subset\mathbb{R}^{2k}$, the following three statements are equivalent:\\
{\it i)} the action $I$ is a harmonic function on $D$;\\
{\it ii)} the complex--valued function $g=S+{\rm i}I$ is analytic on on $D$;\\
{\it iii)} the stationary wavefunction $\psi=\exp(S+{\rm i}I)$ is analytic on $D$.}

The case when the quantum amplitude is spatially constant deserves special attention:

{\bf Property 2} {\sl When $\nabla S=0$ on $D$, the following three statements hold:\\
{\it i)} the action $I$ is a harmonic function on $D$;\\
{\it ii)} the complex--valued function $g$ is analytic on on $D$;\\
{\it iii)} the stationary wavefunction $\psi=\exp(S+{\rm i}I)$ is analytic on $D$.}

\section{The analytic current}

The previous properties follow from an analysis of the conservation law $\nabla\cdot{\bf J}=0$, which holds exactly both clasically and quantum mechanically. 

Let us exchange the variables $S$ and $I$ in the Lagrangian (\ref{eleeseu}). Denoting the result by $\tilde{\cal L}$, we have
\begin{equation}
\tilde{\cal L}=\exp(2I)\left\{{\rm i}\hbar(\partial_tI+{\rm i}\partial_tS)-\frac{\hbar^2}{2m}\left[(\nabla S)^2+(\nabla I)^2\right]-V\right\}.
\label{tildeleeseu}
\end{equation}
The above Lagrangian is manifestly invariant under the transformations $S\rightarrow S+\beta$, where $\beta\in\mathbb{R}$. These transformations induce rigid phase rotations $\tilde\psi=\exp(I+{\rm i}S)\rightarrow\exp({\rm i}\beta)\tilde\psi$ of the wavefunction. The corresponding conserved Noether current is
\begin{equation}
\tilde{\bf J}=\frac{\hbar}{m}\exp(2I)\nabla S.
\label{jota}
\end{equation}
Indeed,
\begin{equation}
\nabla\cdot\tilde{\bf J}=\frac{\hbar}{m}\exp(2I)\left(2\nabla I\cdot\nabla S+\nabla^2S\right),
\label{nighmer}
\end{equation}
and the bracketed term vanishes by virtue of Eq. (\ref{ryhbfb}), after exchanging $S$ and $I$ in the latter.

A remark is in order. The statement {\it the current $\tilde{\bf J}$ is conserved}\/ means {\it conserved under the equations of motion of the Lagrangian $\tilde{\cal L}$}\/; it does {\it not}\/ imply conservation under the motions corresponding to ${\cal L}$. More precisely, the equations of motion of ${\cal L}$ may, but need not, preserve the property $\nabla\cdot\tilde{\bf J}=0$. By the same token, the current ${\bf J}$ is conserved under the Lagrangian ${\cal L}$, but not necessarily under $\tilde{\cal L}$. 

The current $\tilde{\bf J}$ generates scale transformations on the fields of the Lagrangian ${\cal L}$, where the stationary wavefunction is $\psi=\exp(S+{\rm i}I)$. By Eq. (\ref{nighmer}) we have:

{\bf Property 3} {\sl The current $\tilde{\bf J}$ is conserved by the motions corresponding to the Lagrangian ${\cal L}$ if and only if} 
\begin{equation}
\nabla I\cdot \nabla S=-\frac{1}{2}\nabla^2 S.
\label{nalil}
\end{equation}

{\bf Property 4} {\sl  Whenever $S$ is a harmonic function on $D$, the following three statements are equivalent:\\
{\it i)} the divergence $\nabla\cdot\tilde{\bf J}$ vanishes identically on $D$;\\
{\it ii)} the function $g=S+{\rm i}I$ is analytic on $D$;\\
{\it iii)} the stationary wavefunction $\psi=\exp(S+{\rm i}I)$ is analytic on $D$.}

A particular case of the above property occurs when $S$ is spatially constant. When $\nabla S=0$ on $D$, then $\nabla^2S=0$, and the quantum potential (\ref{cuantist}) vanishes identically. Moreover the quantum Hamilton--Jacobi equation (\ref{qhj}) reduces to its classical counterpart, while the continuity equation for the quantum probability fluid, Eq. (\ref{ryman}), reduces to $\nabla^2 I=0$. The two necessary conditions for analyticity of $g=S+{\rm i}I$, namely $\nabla^2S=0$ and $\nabla^2I=0$, are satisfied. 

{\bf Property 5} {\sl  Whenever $S$ is constant on $D$, it holds that $\nabla\cdot\tilde{\bf J}=0$, and the stationary wavefunction $\psi$ is analytic on $D$.}

\section{Discussion}\label{diskku}

We need to identify the quantum effects responsible for the stationary wavefunction $\psi$ generally {\it not}\/ being a complex--analytic map $\psi:\mathbb{C}^{k}\longrightarrow\mathbb{C}$, where $\mathbb{C}^k$ is classical configuration space $\mathbb{R}^{2k}$.

The classical mechanics of a point particle possesses no natural length scale. Indeed, the classical density function of a point particle is a Dirac delta function, which naturally carries the dimensions of an inverse volume. On the contrary, the quantum mechanics of a particle of mass $m$ carries a natural length scale associated, namely the de Broglie wavelength $\lambda_B=\hbar/(mv)$. Quantum density distributions are usually not sharply localised in space.  Instead of being a Dirac delta, the real part of the function $g=S+{\rm i}I$ in  Eq. (\ref{geje}) is spread out, and a necessary requirement of analyticity (that the real and imaginary parts of $g$ be harmonic functions) need not be satisfied. As a consequence, the quantum wavefunction may, but need not always be, analytic.

These conclusions can be neatly reexpressed through the introduction of a new current $\tilde{\bf J}$, defined in Eq. (\ref{jota}), and the corresponding divergence $\nabla\cdot\tilde{\bf J}$. The new current $\tilde{\bf J}$ is obtained from the standard probability density current ${\bf J}$ by the exchange of $S$ and $I$. We have established necessary, sufficient, and necessary and sufficient conditions that relate the (non)vanishing divergence $\nabla\cdot\tilde{\bf J}$ to the (non)analyticity of the function $g=S+{\rm i}I$ and of the stationary wavefunction $\psi=\exp(g)$. 

We have refrained from calling the nonvanishing divergence $\nabla\cdot\tilde{\bf J}$ {\it an anomaly}\/. Our situation does not exactly match the textbook definition of an anomaly \cite{WEINBERG}, in the sense that we are not always dealing with a classical symmetry that breaks down at the quantum level. Classical mechanics need not always be complex--analytic ({\it e.g.}\/, when configuration space is odd--dimensional), nor need $\nabla\cdot\tilde{\bf J}\neq 0$ always hold in the quantum theory. Still, let us temporarily accept calling a nonvanishing value of $\nabla\cdot\tilde{\bf J}$ {\it an anomaly}\/. Then the adjectives {\it holomorphic}\/ and {\it analytic}\/ come to mind. Now calling our nonvanishing divergence $\nabla\cdot\tilde{\bf J}$ {\it the analytic anomaly}\/ might cause confusion with the well--established {\it holomorphic anomaly}\/ of string theory. Indeed, topological string theory \cite {MARINO} and the holomorphic anomaly have been used in ref. \cite{CODESIDO} to analyse the WKB expansion of quantum mechanics. Altogether, calling $\nabla\cdot\tilde{\bf J}$ an {\it analytic divergence}\/ seems more appropriate.

The relation exhibited in ref. \cite{CODESIDO} between quantum mechanics and topological theories \cite{NASH, SCHWARZ} raises an interesting question \cite{NOI}: could it be that quantum mechanics arises as some kind of {\it topological sector}\/  of some underlying theory?

\vskip.5cm
\noindent
{\bf Acknowledgements} Research supported by grant no. ENE2015-71333-R (Spain).

\end{document}